\documentclass[11pt]{article}

\input epsf.sty

\def\lromn#1{\uppercase\expandafter{\romannumeral#1}}

\usepackage{graphicx,amsmath,amssymb}
\def\lromn#1{\uppercase\expandafter{\romannumeral#1}}

\begin{document}

\vspace{2cm}
\begin{center}
\begin{Large}
{\bf Parity violation 
in radiative emission of neutrino pair from 
metastable states of heavy alkaline earth atoms
}

\end{Large}

\vspace{2cm}
\begin{large}
M. Yoshimura, 
 N. Sasao$^{\dagger}$, and S. Uetake

Center of Quantum Universe, Faculty of
Science, Okayama University \\
Tsushima-naka 3-1-1 Kita-ku Okayama
700-8530 Japan

$^{\dagger}$
Research Core for Extreme Quantum World,
Okayama University \\
Tsushima-naka 3-1-1 Kita-ku Okayama
700-8530 Japan \\
\end{large}
\end{center}

\vspace{4cm}

\begin{center}
\begin{Large}
{\bf ABSTRACT}
\end{Large}
\end{center}

Macro-coherent atomic de-excitation involving a neutrino pair emission, 
radiative emission of neutrino pair
(RENP) $|e\rangle \rightarrow |g\rangle + \gamma +\nu_i \nu_j$
(with $\gamma$ a photon and $\nu_i$ a neutrino
mass eigenstate),
is a new experimental tool
to determine undetermined neutrino
parameters such as
the smallest neutrino mass and
distinction of Majorana and Dirac neutrinos.
The best way to prove
that the atomic RENP process accompanied by
unseen neutrino pair involves 
weak interaction is to measure 
parity violating (PV) quantities.
We quantitatively study how this is achieved. 
The basic mechanism of how a favorable situation for PV may
arise from the fundamental electroweak theory
(extended to incorporate finite neutrino masses)
is emphasized and calculation of dependence of
PV observables on applied magnetic field
is worked out for heavy target atoms
of alkaline earth like level structure such as
Sr, Yb, Hg, Xe.
Numerically calculated  parity violating
rates and asymmetry are presented for  
Yb $J=2 \rightarrow 0$ and $J=0 \rightarrow 0$ RENP.

\vspace{4cm}

Key words

Neutrino mass,
Parity violation, Majorana particle,
Beyond the standard gauge theory

\newpage
\lromn1 
{\bf Introduction}

\vspace{0.3cm}
In remarkable achievements
neutrino oscillation experiments
have succeeded in determining five elements of
the neutrino mass matrix \cite{nu oscillation data};
three mixing angles and two
mass squared differences.
They however left
undetermined  important parameters, possibly three parameters
in the Majorana neutrino case; 
the absolute neutrino mass (or the smallest neutrino mass)
scale and the two Majorana CP phases.
Conventional targets  in ongoing experiments of exploring 
a part of these
undetermined neutrino parameters have been nuclei.
Direct measurement of the end point
spectrum of beta decay such as tritium \cite{tritium} and 
(neutrino-less) double beta
decay \cite{nu0 beta} are two main methods to resolve
these outstanding problems.
One serious problem of nuclear target experiments is
the remoteness of released nuclear energies
from the  expected small neutrino mass of a fraction of eV.

In a series of theoretical papers
we proposed and elaborated a new, systematic
experimental method to probe the neutrino mass matrix
using macro-coherent atomic process, namely
radiative emission of neutrino pair (RENP),
$|e \rangle \rightarrow |g\rangle + \gamma +\nu \nu$
\cite{my-prd-07}, \cite{ptep overview}.
We discussed how to enhance otherwise small 
neutrino pair emission rates 
\cite{yst pra}, \cite{ptep overview},
and how to extract neutrino parameters
from the photon energy spectrum \cite{dpsty-plb}.
In the most recent work
we pointed out how to obtain a much larger RENP
rate \cite{ys-13} using a coherent neutrino
pair emission from the zero-th
component of vector current
much like the enhanced admixture
of different parity states in heavy
atoms \cite{bouchiat 2} for atomic parity violation experiments 
\cite{bouchiat exp}, \cite{commins},
\cite{wieman}.
Our experimental efforts towards RENP 
are briefly described in \cite{ptep overview}.

Parity violation (PV) is one of the most important features
that characterizes weak interaction distinguishing from other interactions.
This feature is automatically built in the standard electroweak theory
as a preference of handedness or chirality of neutrino, 
describing neutrinos by two component spinors
unlike all other charged leptons and quarks of four components.
It has been the key feature to prove
the nature of weak processes since the classic work
of Lee and Yang.
In particular,
parity violation in electron interaction with nuclei
caused by Z-boson exchange 
has been discovered in electron scattering 
and atomic parity violation experiments, thereby
establishing the electroweak unification.

In the present work we examine how parity violation in
our proposed process of RENP may arise
from the fundamental electroweak theory.
If parity violating effects turn out to be large,
it should help to experimentally identify RENP and  prove that
weak interaction is involved in the process.
It would be the best way to reject QED backgrounds
in RENP experiments.
A large parity violating (PV) observable and
asymmetry is expected in the case
of de-excitation between states of different parities,
and we shall work out RENP rates
for alkaline earth like atoms,  examples being (1)  
$(6s 6p) {}^3 P_2$ orbital state at $\sim 2.44$eV of neutral Yb,
(2) ${}^3P_0$ state slightly below this level,
and similar ones for Sr, Hg and Xe, all having
low lying excited states made of  $ns, n'p$
or rare earth atoms of electron-hole pair
of the same quantum numbers.

The rest of this work is organized as follows.
In Section \lromn2 how PV observables in RENP process
may arise in the standard electroweak theory
(with finite neutrino masses) by listing all terms of
neutrino interaction with electrons and quarks
to the leading and next leading orders of 1/mass.
Some technical details on the phase space integral
of neutrino pair variables (helcities and momenta) that have
a direct relevance to emergence of
parity odd quantities are relegated to Appendix B.
It is found that without external electric field
transitions between different parity states
are required for large PV effects.
In Section \lromn3 we show that the best target atoms
are alkaline atoms of two-electron system
which have two metastable states of $^3P_2, ^3P_0$.
To produce large PV effects it is
important to use heavy atoms where
$LS$ coupling scheme breaks down, requiring 
calculations in the intermediate coupling scheme..
Some aspects of the intermediate coupling scheme
for heavy atoms are explained in Appendix C.
In Section \lromn4 basic diagrams and corresponding
perturbation amplitudes are identified.
Importance of hyperfine interaction is stressed.
We then calculate in Section \lromn5 amplitudes of RENP,
emphasizing how the magnetic field dependence
is disentangled.
In Section \lromn6 RENP rates, both parity
conserving (PC) and PV, are calculated.
We then illustrate results of numerical computations
on PV observables; PV asymmetry under field and
circular polarization reversal,
taking the example of the Yb $J=2 \rightarrow 0$ transition.
The section \lromn7 is devoted to Yb $J=0 \rightarrow 0$ 
RENP.
The rest consists of summary and Appendices which
collect important technical details not fully explained
in the main text..

We shall calculate amplitudes using perturbation theory 
in non-relativistic quantum mechanics, hence
the time ordering in higher orders
of perturbation should be treated with care.

Throughout this work we use the natural unit
of $\hbar = c = 1$.

\vspace{0.5cm}
\lromn2 
{\bf How parity odd observables arise in RENP}

\vspace{0.3cm}
Typical RENP experiments use four lasers
for trigger and excitation.
For instance, two continuous wave (CW) lasers
of different frequencies, 
$\omega_i\,, \omega_1 +\omega_2 = \epsilon_{eg}$
with $\epsilon_{eg}$ the energy difference between the initial $|e\rangle$ state and 
the final $|g\rangle$ state,
are used as triggers in counter propagating directions (taken along z-axis),
while two excitation lasers of Raman type of frequencies,
$\epsilon_{p}\,,\omega_{s}$ with  $\epsilon_{p}-\omega_{s} = \epsilon_{eg}$
are irradiated in pulses.
Measured variables at the time of excitation pulse irradiation
are the number of events at each trigger frequency $\omega$.
By repeating measurements at different trigger frequency combinations, one obtains
the photon energy spectrum at different frequency $\omega
= \omega_1 (< \omega_2)$ accompanying the neutrino pair.

The macro-coherent three-body RENP process 
$|e \rangle \rightarrow |g\rangle + \gamma +\nu \nu$
conserves both the energy and the momentum \cite{ptep overview},
giving continuous photon energy spectrum with thresholds.
Note that the spontaneous decay of dipole transition
from atoms conserves the energy alone, hence their spectrum 
is continuous despite of a single particle decay.
In RENP there are six photon energy thresholds at
$\omega_{ij} = \epsilon_{eg}/2 - (m_i+ m_j)^2/2\epsilon_{eg}$
where $m_{i(j)}\,,i,j = 1,2,3$ is  neutrino mass of eigenstate. 
Decomposition into six different threshold regions 
is made possible by excellent energy resolution of 
trigger laser frequencies.

Measurement of the photon energy spectrum is
regarded as a parity even observable 
and arises from parity conserving parts of basic interaction of the weak process. 
The only handle for this experiment is the frequency of trigger lasers, 
whose accuracy of $10^{-15}$ is readily obtained with a rapid development of laser technologies \cite{JAlnis:PRA2008}. 
However, there are many other experimental handles in atomic experiments, giving rise to
measurements of PV quantities with relative easiness.

Identification of parity violating (PV) quantities in RENP
under the circumstance of accompanying unseen neutrino pair
is a non-trivial problem,
and we shall describe our method of how 
identification of PV effects and search for
candidate atoms is made in general terms first,
postponing the choice of target states later.

PV effects arise from interference of two RENP amplitudes
of parity even (PE) and parity odd (PO). 
Note that the rate arising from the squared
PO and the squared PE amplitudes give  PC rates.
Extracting explicitly neutrino pair emission vertex,
interference term may arise in three ways: 
the first way via interference
between one term in $A_0 \nu_1^{\dagger} \nu_2$, and the other
in $\vec{A}\cdot\nu_1^{\dagger} \vec{\sigma}\nu_2  $,
and the second and the third ways between 
two decomposed terms either in $A_0^i$ or $\vec{A}^i\,, i=1,2 \cdots$.
Each of $A_{\alpha}^i, \alpha = 0, 1,2,3$ contains product of 
atomic matrix elements, couplings
and energy denominators in perturbation theory.
We use two component notation for electron operators in $A_{\alpha}$,
following 
the $\gamma_5$-diagonal representation of \cite{my-prd-07}.
Relevant leading terms for PO and PE terms are taken from Appendix A:
written in terms of the electron field operator in the
non-relativistic limit of $\gamma_5-$diagonal 
representation, they are
\begin{eqnarray}
&&
A_0 \propto e^{\dagger}\left(b_{12}
+ \delta_{12}
2 \sin^2\theta_w \vec{\sigma}\cdot \frac{\vec{p}}{m_e}
 + O(\frac{1}{m_e^2})\right) e
+\delta_{12}  j_q^0
\,,\hspace{0.5cm}
j_q^0 = -\frac{1}{2} j_n^0 + \frac{1}{2}(1-4\sin^2 \theta_w) j_p^0
\,,
\\ &&
\vec{A} \propto e^{\dagger}
\left(a_{12} \vec{\sigma}  
+\delta_{12} 2 \sin^2\theta_w \frac{1}{m_e} (\vec{p} - i \vec{\sigma} \times \vec{p}) + O(\frac{1}{m_e^2}) \right) e
\,,
\end{eqnarray}
where couplings $a_{12}, b_{12} $ are of order unity and
written in terms of the neutrino mixing angles.
$m_e$ is the electron mass.
Their explicit forms are given in equations of Appendix A.
The term $j_q^0$ is the nuclear mono-pole current contribution
which gives rise to coherently added constituent numbers \cite{ys-13}.
We disregarded terms of orders
of $1/m_e^2$ and $1/m_N$ (1/ nucleon mass),

In order to calculate rates, both parity conserving (PC) and 
parity violating (PV),
added amplitudes are squared, and one proceeds to
calculate summation over neutrino helicities and
momenta, since neutrino
variables are impossible to measure
under usual circumstances.
Thus,
one deals with a phase space integral of
neutrino pair momenta after helicity summation
in the form,
\begin{eqnarray}
&&
\int d{\cal P}_{\nu} \sum_{h_i} |A_0 \nu_1^{\dagger} \nu_2 + 
\vec{A}\cdot\nu_1^{\dagger} \vec{\sigma}\nu_2 |^2 
\,, \hspace{0.5cm}
 d{\cal P}_{\nu} = \frac{d^3p_1 d^3p_2}{(2\pi)^2}\delta (\omega + E_1 + E_2 - \epsilon_{eg}) 
\delta (\vec{k} + \vec{p}_1 + \vec{p}_2) 
 \,.
\end{eqnarray}
All necessary phase space integrals are listed in
Appendix B.
The non-trivial part of the phase space integral relevant to
PV interference arises from the term 
$-2 \Re (A_0 \vec{A}^*)$ multiplied by
\begin{eqnarray}
&&
\int d{\cal P}_{\nu} ( \frac{\vec{p}_1}{ E_1} + \frac{\vec{p}_2}{ E_2}) = 
\vec{k}  \frac{J_{12}(\omega)}{\omega}
\,, 
\end{eqnarray}
where $J_{12}(\omega)$ is a scalar function given in Appendix B.
The photon momentum vector $\vec{k}$ is thus
multiplied to $-2 \Re (A_0 \vec{A}^*)$,
which give three types of electron operators proportional to
\begin{eqnarray}
&&
\vec{k}\cdot\vec{\sigma}
\,, \hspace{0.5cm} 
\frac{\vec{k}\cdot\vec{p}}{m_e}
\,, \hspace{0.5cm}
 i \frac{\vec{k}\cdot \vec{\sigma}
\times \vec{p} }{m_e}
\,.
\label{pv candidates}
\end{eqnarray}
All of these are hermitian.
The only PO operator is the first one, $\vec{k}\cdot\vec{\sigma}$.
The remaining amplitude multiplied to this involves PC QED interaction such
as $\vec{d}\cdot \vec{E}$, hence this term alone can
be adopted for PO amplitude.

The fact that PV term arises without the suppression of $1/m_e$
might appear surprising.
This conclusion is consistent with
the ordinary view that PV effects must arise from interference
of parity odd combination of $V\cdot A$.
The spin current of electron $\propto \vec{\sigma}$ arises from
spatial component of 4-axial vector $\propto \gamma^{\alpha}\gamma_5$
in the non-relativistic limit,
while the nuclear mono-pole current $\propto j_q^0$
arises from time component of 4-vector current $\propto \gamma_{\alpha}$.
It is the unique combination of electron and nuclear current
operators 
that gives rise to large PV terms without the suppression of
$1/m_e$ order, which became possible only with the advent
of nuclear mono-pole contribution given in \cite{ys-13}.
This excludes the other two possibilities of having $1/m_e$
suppression for PV effect.
We refer to Appendix A on technical aspects of
PO operators.

\vspace{0.5cm}
\lromn3 
{\bf Basic mechanism in heavy alkaline earth atoms}

\vspace{0.3cm}
We shall first examine consequences of the conclusion of the previous section that 
RENP transition for a large PV measurement 
involves states of different parities in the initial and the final
states.
The simplest possibility might be use of the lowest excited
state of alkali atoms taken as the initial $|e \rangle$ state.
The RENP of alkali atom however must compete with a fast E1 transition,
and it may be experimentally difficult to measure PV  quantity 
under a large signal to the background ratio.
We shall not consider this possibility any further.

The next easiest may be two-electron system consisting of
angular momentum combination of parity odd $sp$ (two-electron system of the
orbital angular momentum $l=0$ and $1$).
This combination appears as the first excited group of levels
in alkaline earth atoms.
Two electrons may be either in the spin triplet or the spin
singlet state in the terminology of the $LS$ coupling scheme.
Thus, one has four different states (with the usual magnetic
degeneracy of energies), ${}^3P_2, {}^3P_1, {}^3P_0, {}^1P_1$,
the atomic term symbol of ${}^{2S+1}L_J$ being used \cite{atomic physics}.

Another important consideration is that it is better
to use heavy (large atomic number) atoms for
large RENP rates \cite{ys-13}.
This poses a problem of state mixing in the $LS$ scheme,
which requires the use of intermediate coupling scheme
\cite{condon-shortley}.
The $LS$ coupling scheme is based on
the assumption that electrostatic interaction between electrons is
much larger than the spin-orbit interaction
$\sum_i \xi(r_i) \vec{l}_i\cdot\vec{s}_i$.
In heavy atoms such as Pb, however, the spin-orbit interaction becomes larger and the $jj$ coupling scheme becomes a better description 
\cite{condon-shortley}. Nevertheless most of heavy atoms is well described by the intermediate coupling scheme using the
$LS$ basis.

In the intermediate coupling scheme applied to heavy
alkaline earth atoms, one considers the mixing among states of
the same total angular momentum.
This is because the total 
angular momentum is conserved under the presence of
the spin-orbit interaction.
This type of mixing occurs for ${}^3P_1$ 
and ${}^1P_1$ of the $LS $ scheme.
Energy eigenstates are given in terms of the $LS$ basis,
\begin{eqnarray}
&&
|{}^+P_1 \rangle = \cos \theta |{}^1P_1 \rangle 
+ \sin \theta |{}^3P_1 \rangle
\,, \hspace{0.5cm}
|{}^-P_1 \rangle = \cos \theta |{}^3P_1 \rangle 
- \sin \theta |{}^1P_1 \rangle
\,,
\end{eqnarray}
(with $\pm$ denoting larger/smaller energy state)
where the angle $\theta$ is determined by
the strength of spin-orbit interaction in the system
and is related to experimental data
of level energies.
In the Yb case $\sin \theta \sim 0.16$,
as shown in Appendix D, where some further details of the intermediate coupling schem are also described.
A relatively large
dipole moments $d(|{}^-P_1 \rangle \rightarrow |{}^1S_0\rangle)$ 
needed for sizable RENP 
is induced by a non-vanishing value of $\theta$.

\begin{figure*}[htbp]
 \begin{center}
 \epsfxsize=0.6\textwidth
 \centerline{\epsfbox{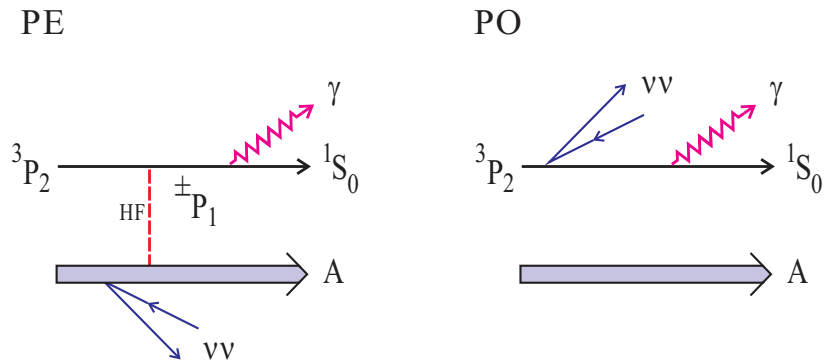}} \hspace*{\fill}
   \caption{Parity even (PE) and odd (PO) 
diagram contributions. HF given by dashed line is
hyperfine mixing interaction as described in the text.
 }
   \label {yb pe-po diagrams}
 \end{center} 
\end{figure*}

We shall consider ${}^3P_2$ de-excitation for RENP.
The ${}^3P_0$ case is treated separately in Section \lromn7.
Due to the quantum number changes both in the orbital
and the spin parts one needs at least two
virtual transitions for de-excitation of ${}^3P_2 \rightarrow {}^1S_0$.
Relevant PE and PO diagrams leading to large amplitudes
are depicted in Fig(\ref{yb pe-po diagrams}).
PO diagram involves valence electron alone and
contains matrix elements of the spin current of
neutrino pair emission and E1 photon emission.
PE diagram contains a large nuclear mono-pole current
assisted by hyperfine interaction, which causes
the necessary quantum number change to valence
electron.

Hyperfine interaction is caused by the nuclear magnetic
field and quadrupole field.
For simplicity we shall consider the magnetic
hyperfine interaction alone by taking the nuclear
spin of $I=1/2$ which excludes the possibility of
quadrupole interaction \cite{weissbluth}.
Both Yb and Xe have isotopes of large
natural abundances of this spin magnitude.
The magnetic hyperfine interaction consists of
dipole-dipole interaction and Fermi contact
interaction,
\begin{eqnarray}
&&
{\cal H}_h = \vec{I}\cdot\vec{A}
\,, \hspace{0.5cm}
\vec{A} = g_e g_N \mu_B \mu_N 
\left( \sum_i
\frac{\vec{l}_i-\vec{s}_i}{r_i^3}
+ \frac{3\vec{r}_i \vec{s}_i\cdot\vec{r_i}}{r_i^5}
+ \frac{8\pi}{3} \delta(\vec{r})\vec{s}_i 
\right)
\,.
\label{magnetic hyperfine interaction}
\end{eqnarray}
In the alkaline earth atom both
of these contribute, the p-electron participating
in the dipole-dipole interaction while the s-electron
in the Fermi interaction.
Matrix elements of this interaction are given in
Appendix.

The main effect of hyperfine interaction
in the PE amplitude is to give rise to non-vanishing
vertex between ${}^3P_1$ component in the levels
${}^{\pm}P_1$ and ${}^3P_2$, which is given by
\begin{eqnarray}
&&
\langle {}^1S | \vec{d}\cdot\vec{E}|{}^{\pm}P_1 \rangle
\langle {}^{\pm}P_1| {\cal H}_h | {}^3P_2 \rangle
\,.
\end{eqnarray}
Mixing caused by hyperfine interaction
gives non-vanishing off-diagonal matrix
elements for $F=3/2$ states,
\begin{eqnarray}
&&
\langle {}^-P_1 3/2| \vec{I}\cdot\vec{A} | {}^3P_2 3/2\rangle =
\frac{\sqrt{5}}{8} (b - \frac{6}{5}a) \cos \theta
+ \frac{1}{4}\sqrt{\frac{5}{2}} (b - \frac{8}{5}a) \sin \theta
\,,
\\ &&
\langle {}^+P_1 3/2| \vec{I}\cdot\vec{A} | {}^3P_2 3/2\rangle =
- \frac{1}{4}\sqrt{\frac{5}{2}} (b - \frac{8}{5}a) \cos \theta
+ \frac{\sqrt{5}}{8} (b - \frac{6}{5}a)\sin \theta 
\,,
\\ &&
a = g_e g_N \mu_B \mu_N\langle 6p | \frac{1}{r^3} | 6p \rangle
\,, \hspace{0.5cm}
b = g_e g_N \mu_B \mu_N\frac{8\pi}{3}|\psi_{6s}(0)|^2
\,.
\end{eqnarray}
Hyperfine parameters $b, a$ may be determined by
experimental data of hyperfine splitting of ${}^3P_2, {}^{\pm}P_1$, as
discussed in Appendix.
The dominant parameter for Yb is $b \sim 13 $GHz
and the mixing amplitude is of order
$\sqrt{5}b \cos \theta /8 \sim 3.5$GHz $\sim 2.3 \mu$eV.

\vspace{0.5cm}
\lromn4 
{\bf Magnetic factors and PV observables}

\vspace{0.3cm}
We consider application of external magnetic field
to orient atoms which makes easier to produce
various types of PV observables.
The simplest among these is a PO angular
distribution of emitted photon from polarized atom.
The magnetic field is applied, for generality, to a tilted direction
from the trigger axis (defined by z-axis) by an
angle $\theta_m$.
All projected angular momenta are taken along
the magnetic field direction.
Hence the angular part is described by
\begin{eqnarray}
&&
|J, \tilde{M} \rangle = e^{-i\theta_m J_y} | J, M \rangle
= \sum_{M'} d^J_{M,M'}(\theta_m) | J, M' \rangle
\,,
\end{eqnarray}
where $d^J_{M,M'}(\theta_m)$ is the Wigner d-function 
or the rotation matrix in the terminology of \cite{rose}.
Furthermore, two types of circular polarization, R and L or
$h= \pm 1$, for trigger (hence RENP emitted) photon
are considered.
Amplitudes and rates are thus functions of $\theta_m$
and $h$.
Helicity dependence is readily worked out
by taking the E1 dipole element proportional to
components of spherical harmonics $Y_{1, \pm 1}$.

Components of magnetic factors are defined
by various matrix elements sandwiched between
these tilted states.
For instance, E1 emission is given by
matrix elements of
\begin{eqnarray}
&&
d \langle J, \tilde{M} | Y_{1, \pm 1}|J', \tilde{M}' \rangle = 
d \sum_{M_1, M_2} d^J_{M, M_1} d^{J'}_{M', M_2} 
\langle J, M_1| Y_{1, \pm 1}|J', M_2 \rangle
\nonumber \\ &&
= d \sum_{M_2} d^J_{M, M_2 \pm 1} d^{J'}_{M', M_2} 
\langle J, M_2 \pm 1| Y_{1, \pm 1}|J', M_2 \rangle
\,,
\end{eqnarray}

RENP rates are functions of the angle $\theta_m$
and circular polarization $h= \pm$ of the
trigger field.
Two readily calculable PV asymmetries are
rate differences under the magnetic field
reversal and under the polarization reversal.
We call these two asymmetries as
PV asymmetry under field reversal and
symmetry under polarization reversal.

For definiteness let us consider RENP
transition from ${}^3P_2, F=3/2$ where
$\vec{F}$ is the angular momentum sum of
electrons and nucleus.
Magnetic angular factors we need for PV rate
differences are
\begin{eqnarray}
&&
{\cal M}_B(x)= -\sqrt{3} \cos^3 x
\,, \hspace{0.5cm}
{\cal M}_h(x) =\frac{\sqrt{3}}{4}(1+ 3 \cos (2x))
\,,
\end{eqnarray}
and for PC quantities
\begin{eqnarray}
&&
(1) \; 
{\cal M}_{{\rm PC1}}(x)= \frac{1}{4}(3 + \cos (2x)\,)
\,,
\\ &&
(2) \; 
{\cal M}_{{\rm PC2}}(x) = \frac{3}{4}(2 + \cos (2x) + \cos(4x) \,)
\,.
\end{eqnarray}
PV asymmetry ${\cal M}_B$ is related to the one under
field reversal, while ${\cal M}_h(x) $ to the one under polarization
reversal.
These functions are derived from combinations of
Wigner d-functions in Appendix.
In PV asymmetry under field reversal 
one needs the difference in two directions,
$x$ and $\pi-x$; ${\cal M}_B(x) - {\cal M}_B(\pi-x)
\neq 0$.
The simplest PV asymmetry of this kind is
the forward-backward asymmetry for $x=0$.
On the other hand,
PV asymmetry under polarization reversal requires
an integrated quantity
$\int_{-1}^1 dx {\cal M}_h (x)$.
These functions are divided by PE combinations 
of rates derived from ${\cal M}_{{\rm PC1}}(x), {\cal M}_{{\rm PC2}}(x)$
in order to define normalized asymmetries.
Integrated quantities are denoted by 
$\tilde{{\cal M}}_i = \int_{-1}^1 dx {\cal M}_i (x)$ for subsequent use.

We do not apply external electric field field
to avoid possible confusion under instrumental 
effect \cite{yfsy-10}.

\vspace{0.5cm}
\lromn5
{\bf PC rates and PV asymmetry for ${}^{171}$Yb$\,{}^3P_2$ RENP}

\vspace{0.3cm}
RENP spectral rates may be expressed by two
formulas $\Gamma_{2\nu\gamma}^{\pm}(\omega)$ 
which are interchanged by reversal of instrumental
polarity; the magnetic field direction.
Rates may be written as
\begin{eqnarray}
&&
\Gamma_{2\nu\gamma}^{\pm}(\omega) = \Gamma_{2\nu\gamma}^{PC1}(\omega)
+ \Gamma_{2\nu\gamma}^{PC2}(\omega) \pm \Gamma_{2\nu\gamma}^{PV}(\omega)
\,.
\end{eqnarray}
The last term is the interference term arising from
the product of PE and PO amplitudes, while
the first two terms result from the squared PE and PO amplitudes.
We decompose these three spectral rates, 
both parity conserving (PC) and parity violating (PV),
into an overall factor denoted by $\Gamma_0$, various spectral shape functions
of kinematical nature,
atomic factors, and the dynamical factor $\eta_{\omega}(t)$.
We shall use a unit of 100 MHz for A-coefficients (decay rates)
and eV for all energies.
We give rates appropriate for Yb $J=2 \rightarrow 0$ RENP.
The conversion factor in our natural unit
is $\hbar c = 1.97 \times 10^{-5} {\rm eV}\cdot{\rm cm}$.

The overall RENP rate is given by
\begin{eqnarray}
&&
\Gamma_0 = 
\frac{3}{4} G_F^2 \epsilon_{eg} n^3 V  
\frac{\gamma_{+g}}{\epsilon_{+g}^3}   \eta_{\omega}(t)
\sim 54 {\rm mHz} (\frac{n}{10^{21}{\rm cm}^{-3}})^3 \frac{V}{10^2 {\rm cm}^3} 
\frac{\epsilon_{eg}}{{\rm eV}} 
\frac{\gamma_{+g}{\rm eV}^{3}}{\epsilon_{+g}^3{\rm 100 MHz}} 
 \eta_{\omega}(t)
\,.
\label{overall rate}
\end{eqnarray}

The factor $\eta_{\omega}(t)$ is the extractable fraction of
field intensity $\epsilon_{eg}n$ stored in the initial upper level
$|e\rangle$. The storage and development of
target polarization is induced by two trigger laser irradiation
of $\omega + \omega' = \epsilon(n'p) - \epsilon(ns), \omega < \omega'$.
The storage is due to a second order QED process,
M1$\times$E1 type of two-photon paired super-radiance (PSR), 
in alkaline atoms.
The calculation of $\eta_{\omega}(t)$ requires numerical solution of the master
equation for developing fields and target polarization
given in \cite{yst pra}, \cite{ptep overview}.
Usually, $\eta_{\omega}(t)$ is much less than unity,
and depends on experimental conditions.

Energy denominator factors are given by
\begin{eqnarray}
&&
{\rm PE}; \hspace{0.3cm}
f_0(\omega) =  \frac{\sqrt{5}b}{8(\epsilon_{eg} - \omega)}
\left( \frac{c_+}{\epsilon_{+g} - \omega}+
\frac{\gamma_{-g} \epsilon_{+g}^3}{\gamma_{+g} \epsilon_{-g}^3}\frac{c_-}{\epsilon_{-g} - \omega} 
\right)
\,,
\\ &&
c_+ = (1- \frac{6a}{5b}) \sin \theta - 2\sqrt{2}(1- \frac{8a}{5b}) \cos \theta 
\,,
\\ &&
c_- = (1- \frac{6a}{5b}) \cos \theta + 2\sqrt{2}(1- \frac{8a}{5b}) \sin \theta
\,,
\\ &&
{\rm PO}; \hspace{0.3cm}
f_1(\omega) = - \frac{1}{\epsilon_{+g} - \omega} -
\frac{\gamma_{-g} \epsilon_{+g}^3}{\gamma_{+g} \epsilon_{-g}^3}\frac{1}{\epsilon_{-g} - \omega} 
\,,
\end{eqnarray}
with $b, a$ two hyperfine constants, and $\theta$ the spin-orbit mixing.

Individual contributions of remaining factors are as follows.

(1) Nuclear mono-pole PC rate assisted by hyperfine interaction is given by
\begin{eqnarray}
&&
\Gamma_{2\nu \gamma}^{PC1} = \Gamma_0 f_0^2(\omega)
Q_w^2  
I(\omega)
\frac{3}{4} \tilde{{\cal M}}_{PC1} 
\,, \hspace{0.5cm}
I(\omega) =  2\pi \sum_i I_{ii}(\omega)
\theta(\omega_{ii} - \omega)
\,,
\\ &&
I_{ii}(\omega) = 
\frac{1}{2}(\, C_{ii}(\omega)+ A_{ii}(\omega)
 + \delta_M m_1 m_2 D_{ii}(\omega) 
 \,)
\,, \hspace{0.5cm}
Q_w = N - 0.044 Z
\,.
\end{eqnarray}

(2) 
PC rate arising from squared 
valence spin current contribution is 
\begin{eqnarray}
&&
\Gamma_{2\nu \gamma}^{PC2} = \Gamma_0 f_1^2(\omega)
H(\omega; \theta_m) \frac{1}{4} \tilde{{\cal M}}_{PC2}  
\,, \hspace{0.5cm}
H(\omega; \theta_m) = 2\pi \sum_i a_{ii}^2 H_{ii}(\omega)
\theta(\omega_{ii} - \omega)
\,,
\\ &&
H_{ii}(\omega) = \frac{1}{2} \left(
C_{ii}(\omega) - A_{ii}(\omega) - \delta_M m_i^2 D_{ii}(\omega)
\right) + \frac{B_{ii}(\omega)}{\omega^2}
\,.
\end{eqnarray}

(3) Interference term between PO and PE amplitudes
is given by
\begin{eqnarray}
&&
\Gamma_{2\nu \gamma}^{PV} =  -2\Gamma_0  f_0(\omega)f_1(\omega)
Q_w \frac{\sqrt{3}}{4}\tilde{{\cal M}}_{h}   
\,,
\label{interference rate}
\\ &&
J(\omega) =  2\pi \sum_i a_{ii} J_{ii}(\omega) \theta(\omega_{ii} - \omega)
\,,\hspace{0.5cm}
J_{ii}(\omega) = 
- \frac{\Delta_{ii}(\omega)}{4\pi}\omega
\left(
\epsilon_{eg} -\frac{4}{3} \omega
 +\frac{4 (\epsilon_{eg} - \omega) m_i^2 }{3\epsilon_{eg}  (\epsilon_{eg} -2 \omega) }
\right)
\,.
\end{eqnarray}
We refer to Appendix B for $\Delta_{ii}(\omega)\,, A_{ii}(\omega)\,, B_{ii}(\omega)\,,
C_{ii}(\omega)\,, D_{ii}(\omega)$ that arise from the neutrino phase space
integration.
For simplicity we wrote down formulas under polarization reversal.
Formulas relevant to the field reversal are obtained by
replacing factors $\tilde{{\cal M}}_i $ by differences at $x=0$
and $x=\pi$ of corresponding functions ${\cal M}_i(x)$.

PV asymmetry is defined with appropriate choice
or combinations of $W_i$ factors as discussed
in the preceding section, normalized to
\begin{eqnarray}
&&
{\cal A}(\omega) = \frac{2\Gamma_{2\nu \gamma}^{PV} }
{\Gamma_{2\nu \gamma}^{PC1} + \Gamma_{2\nu \gamma}^{PC2}}
\,.
\label{pv-asymmetry}
\end{eqnarray}
This is a quantity to be compared with
the experimental asymmetry obtained by
taking the ratio of the difference to the sum of two
rates under the field or the polarization reversal.

\vspace{0.5cm}
\lromn6 
{\bf Numerical calculation of RENP spectral rates}

\vspace{0.3cm}
We numerically computed spectral rates for Yb and Xe
atoms of ${}^3P$ states.
Below we discuss and show results of Yb, since
it gives larger rates.

A-coefficients we need for computations of Yb RENP
are 
$\gamma_{+g} = 176, \gamma_{-g} =1.1$MHz's and
$\epsilon_{+g} =3.108, \epsilon_{-g} =2.2307, 
\epsilon({}^3P_2) = 2.4438
$eV's.
In $^{171}$Yb $^3P_2$ RENP,
the dominance of the intermediate state ${}^+P_1$ is evident:
$f_0(\omega) \propto - 2.6/(\epsilon_{+g}-\omega) + 0.0024/(\epsilon_{-g}-\omega)$.

Hyperfine split energies we use for ${}^{171}$Yb
parameter determination (of $b, a$) are
\cite{TWakui:JPSJ2003}, \cite{DLClark:PRA1979},
\cite{DDas:PRA2005},
\begin{eqnarray}
&&
\epsilon({}^3P_2\, 5/2) - \epsilon({}^3P_2\, 3/2) = 6.68 {\rm GHz}
\,, \hspace{0.5cm}
\epsilon({}^-P_2\, 3/2) - \epsilon({}^-P_2\, 1/2) = 5.94 {\rm GHz}
\,, 
\nonumber \\ &&
\epsilon({}^+P_2\, 3/2) - \epsilon({}^+P_2\, 1/2) = - 0.32 {\rm GHz}
\,.
\end{eqnarray}
These give hyperfine parameters of $ {}^{171}$Yb$\,{}^3P_2$,
$b \sim 13, a \sim 0.17$GHz's, as discussed in Appendix D.
We use these hyperfine constants and the spin-orbit mixing
$\theta$ as determined by energy levels of 
${}^3P_2, {}^{\pm}P_1, {}^3P_0$
in Appendix C.

\begin{figure*}[htbp]
 \begin{center}
 \epsfxsize=0.6\textwidth
 \centerline{\epsfbox{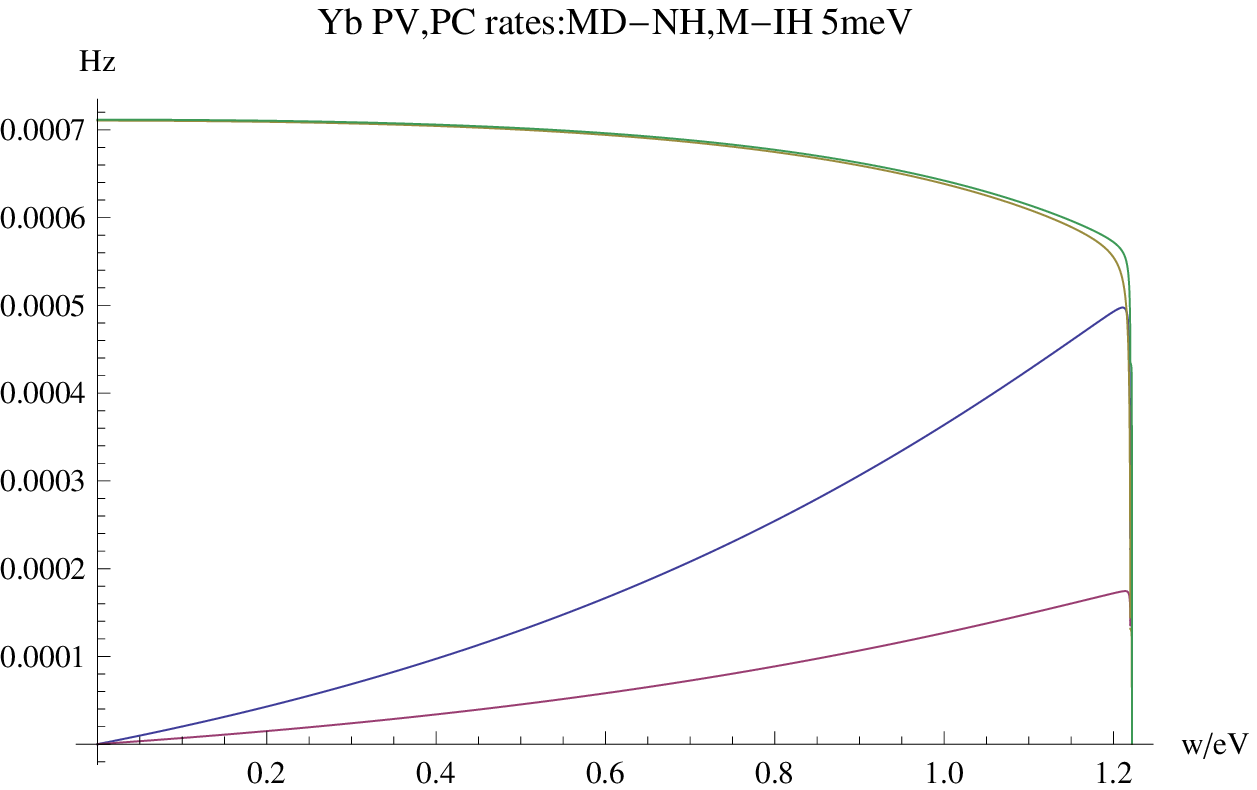}} \hspace*{\fill}
   \caption{${}^{171}$Yb RENP PC rates, and PV rate differences
   under field reversal 
for Majorana neutrino pair emission of smallest mass 5meV,
   PV-NH in blue andPV-IH in  magenda,
   PC-NH in brown and PC-IH in green.
   PC rates are scaled down by 1/500 for easy comparison.
Assumed parameters are
   target number density $=10^{22}$cm$^{-3}$,
   target volume $=10^2$cm$^3$. $\eta_{\omega}(t) = 1$ is taken here and
   in all following figures.
}
   \label{pvc mnih overall}
 \end{center} 
\end{figure*}

\begin{figure*}[htbp]
 \begin{center}
 \epsfxsize=0.6\textwidth
 \centerline{\epsfbox{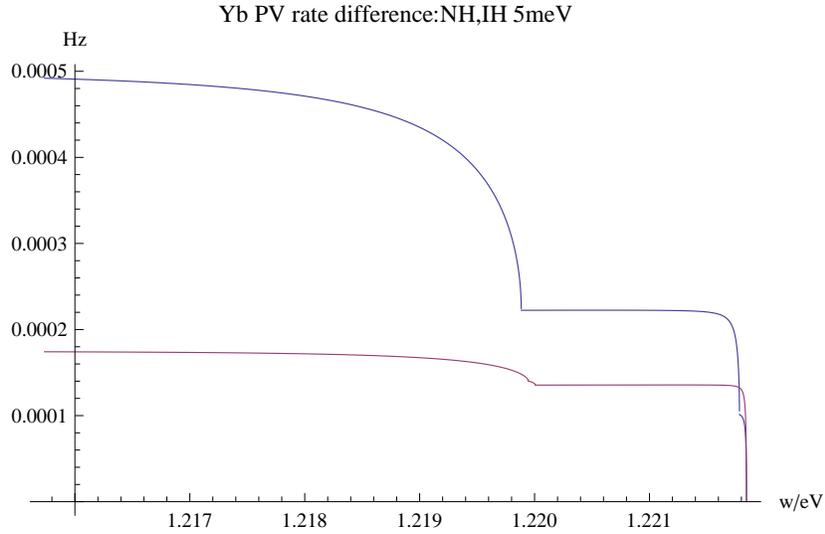}} \hspace*{\fill}
   \caption{${}^{171}$Yb PV rates in the threshold region
corresponding to Fig(\ref{pvc mnih overall}).
NH in blue and IH in magenda.
 }
   \label{pv mnih th}
 \end{center} 
\end{figure*}

\begin{figure*}[htbp]
 \begin{center}
 \epsfxsize=0.6\textwidth
 \centerline{\epsfbox{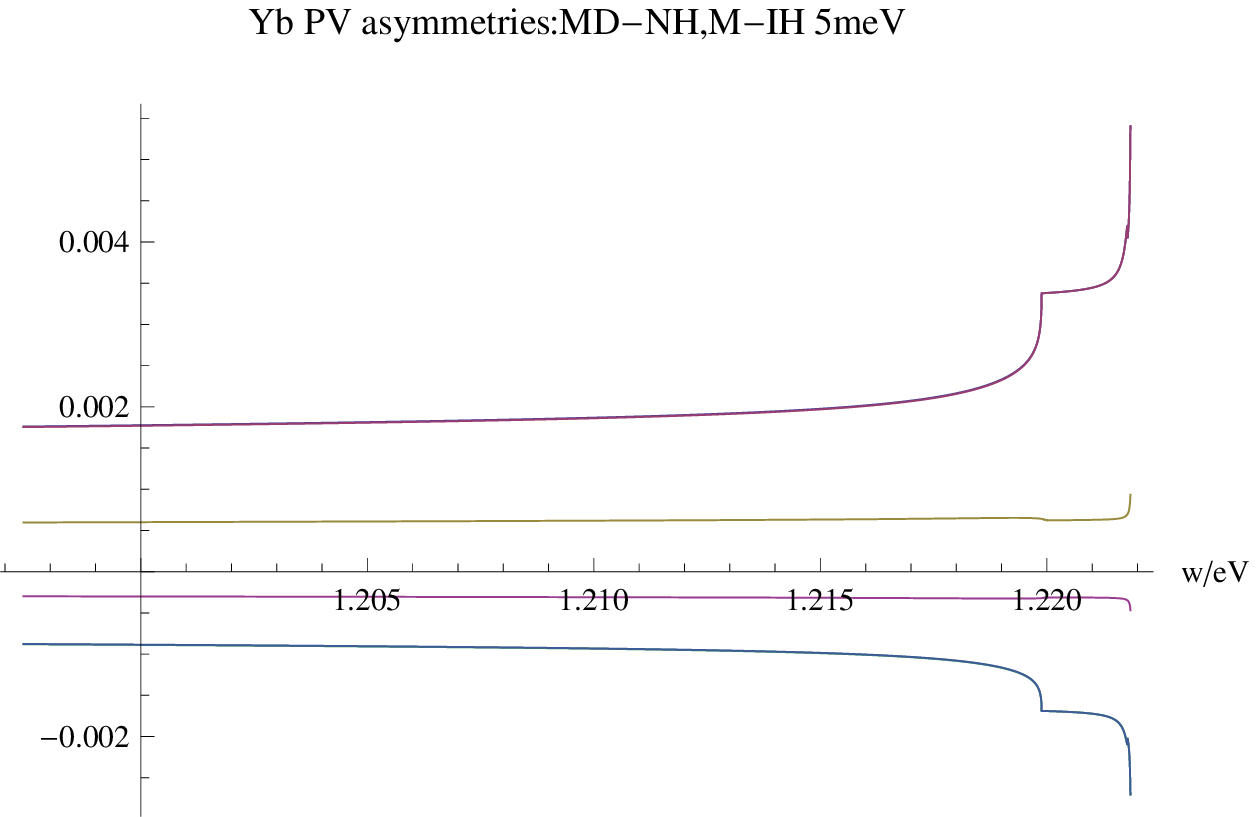}} \hspace*{\fill}
   \caption{${}^{171}$Yb PV asymmetries under field reversal
and polarization reversal.
M-NH in blue, D-NH in magenda and M-IH in brown
in the positive side for field reversal, the forward-backward asymmetry.
The negative value side is for PV asymmetry under polarization reversal;
M-NH in green, D-NH in blue, and M-IH in magenda.
MD differences are difficult to resolve with this resolution.
 }
   \label{pv asym 2}
 \end{center} 
\end{figure*}

PC rates and PV rate differences under field
and polarization reversals are illustrated for the smallest mass of 5 meV
in Fig(\ref{pvc mnih overall})
$\sim$ Fig(\ref{pv asym 2}).
In these figures, N Hz of rates means N number of
events per second.
It is difficult to distinguish the Majorana case
from the Dirac case in absolute rates and PV rate
differences, as seen in Fig(\ref{pvc mnih overall})
and Fig(\ref{pv mnih th}).
But it is possible to compare Majorana-Dirac differences
from PV asymmetries in lower photon energies, assuming that
one can obtain a large statistics data,
as seen in Fig(\ref{pv asym 2}).
It is difficult to distinguish the Majorana neutrino pair emission from
the Dirac pair emission in the examined cases.

\vspace{0.5cm}
\lromn7 
{\bf ${}^{171}$Yb$\,{}^3P_0$ RENP}

\vspace{0.3cm}
Finally, we discuss the case of $0\rightarrow 0$ transition.
The process is of special interest, because the single photon
emission is highly forbidden for this case.
It turns out that PV observables are more restricted than
${}^3P_2$  RENP.
We describe main results briefly, since the method of computations
is already explained in the ${}^3P_2$ case,

Contributions from Fermi contact interaction
and dipole-dipole interaction are calculated
by taking explicit forms of angular parts of
wave functions derived by addition of angular momenta.
Results of hyperfine mixing are
\begin{eqnarray}
&&
\langle {}^3P_1\, F=1/2|  \vec{I}\cdot\vec{A}
| {}^3P_0\,  F=1/2 \rangle = - \frac{1}{2\sqrt{2}} (b- 2a)
\,,
\\ &&
\langle {}^1P_1\, F=1/2|  \vec{I}\cdot\vec{A}
| {}^3P_0 \, F=1/2 \rangle = - \frac{1}{4} (b+ 2a)
\,,
\\ &&
\langle {}^- P_1\, F=1/2|  \vec{I}\cdot\vec{A}
| {}^3P_0\,  F=1/2 \rangle =
- \frac{1}{2\sqrt{2}} (b- 2a) \cos \theta
+  \frac{1}{4} (b+ 2a)\sin \theta
\,,
\\ &&
\langle {}^+ P_1\, F=1/2|  \vec{I}\cdot\vec{A}
| {}^3P_0 \, F=1/2 \rangle =
-  \frac{1}{4} (b+ 2a)\cos  \theta
- \frac{1}{2\sqrt{2}} (b- 2a) \sin \theta
\,.
\end{eqnarray}

Since both initial and final states have angular momentum
$J=0$, effects of tilted magnetic field are much simplified
than the ${}^3P_2$ case.
PV rate differences and PC rates have the following
angular dependences:
\begin{eqnarray}
&&
{\rm PV} ; \hspace{0.3cm}
W_3^{\pm }(x)W_4^{\pm }(x) = \frac{1}{2} \sin^2 x \cos x
\,,
\\ &&
{\rm PC1} ; \hspace{0.3cm}
(W_3^{\pm }(x)\,)^2 = \frac{1}{2} \sin^2 x 
\,,
\\ &&
{\rm PC2} ; \hspace{0.3cm}
(W_4^{\pm }(x)\,)^2 = \frac{1}{2} \sin^2 x \cos^2 x
\,.
\end{eqnarray}
From these we conclude that PV asymmetry under
polarization reversal given by angular integrated
quantities vanish, while PV asymmetry under
field reversal $x \rightarrow \pi -x$ is
non-vanishing, with
\begin{eqnarray}
&&
{\rm PV\; asymmetry}\;
\propto
\sum_{\pm} \left(
W_3^{\pm }(x)W_4^{\pm }(x) - W_3^{\pm }(\pi - x)W_4^{\pm }(\pi - x)
\right) = \sin^2 x \cos x
\,.
\end{eqnarray}
Note that this quantity vanishes at $x=0$.

We now turn to rate formulas.
The overall RENP rate is the same as in the previous 
${}^3P_2$ case, while
energy denominator factors are given by
\begin{eqnarray}
&&
{\rm PE}; \hspace{0.3cm}
f_0(\omega) = \frac{b}{4(\epsilon_{eg} - \omega)} 
(\frac{c_+^{(0)}}{\epsilon_{+g} - \omega}+
\frac{\gamma_{-g} \epsilon_{+g}^3}{\gamma_{+g} \epsilon_{-g}^3}
\frac{c_-^{(0)}}{\epsilon_{-g} - \omega} )
\,,
\\ &&
c_+^{(0)} 
= \sqrt{2}(1- \frac{2a}{b}) \sin \theta + (1+ \frac{2a}{b}) \cos \theta 
\,,
\\ &&
c_-^{(0)}
 = \sqrt{2}(1- \frac{2a}{b})) \cos \theta  - (1+ \frac{2a}{b}) \sin \theta
\,,
\\ &&
{\rm PO}; \hspace{0.3cm}
f_1(\omega) =  \sin \theta \cos \theta
(\frac{1}{\epsilon_{+g} - \omega}+
\frac{\gamma_{-g} \epsilon_{+g}^3}{\gamma_{+g} \epsilon_{-g}^3}\frac{1}{\epsilon_{-g} - \omega} )
\end{eqnarray}

We thus derive individual contributions of remaining factors in the following.

(1) Nuclear mono-pole PC rate assisted by hyperfine interaction is given by
\begin{eqnarray}
&&
\Gamma_{2\nu \gamma}^{PC1} = \Gamma_0
Q_w^2f_0^2(\omega)
I(\omega)
\frac{1}{16} ( W_3^{\pm})^2
\,, \hspace{0.5cm}
I(\omega) =  2\pi \sum_i I_{ii}(\omega)
\theta(\omega_{ii} - \omega)
\,,
\\ &&
I_{ii}(\omega) = 
\frac{1}{2}(\, C_{ii}(\omega)+ A_{ii}(\omega)
 + \delta_M m_1 m_2 D_{ii}(\omega) 
 \,)
\,, \hspace{0.5cm}
Q_w = N - 0.044 Z
\,.
\end{eqnarray}

(2) 
PC rate arising from squared 
valence spin current contribution is 
\begin{eqnarray}
&&
\Gamma_{2\nu \gamma}^{PC2} = \Gamma_0 f_1^2(\omega)
H(\omega; \theta_m) 
\frac{2}{3} ( W_4^{\pm})^2
\,, \hspace{0.5cm}
H(\omega; \theta_m) = 2\pi \sum_i a_{ii}^2 H_{ii}(\omega)
\theta(\omega_{ii} - \omega)
\,,
\\ &&
H_{ii}(\omega) = \frac{1}{2} \left(
C_{ii}(\omega) - A_{ii}(\omega) - \delta_M m_i^2 D_{ii}(\omega)
\right) + \frac{B_{ii}(\omega)}{\omega^2}
\,.
\end{eqnarray}

(3) Interference term between PO and PE amplitudes
is given by
\begin{eqnarray}
&&
\Gamma_{2\nu \gamma}^{PV} =  -2\Gamma_0  f_0(\omega) f_1(\omega)
Q_w \frac{\sqrt{2}}{4 \sqrt{3}} W_1^{\pm}  W_2^{\pm}
\,,
\label{interference rate}
\\ &&
J(\omega) =  2\pi \sum_i a_{ii} J_{ii}(\omega) \theta(\omega_{ii} - \omega)
\,,\hspace{0.5cm}
J_{ii}(\omega) = 
- \frac{\Delta_{ii}(\omega)}{4\pi}\omega
\left(
\epsilon_{eg} -\frac{4}{3} \omega
 +\frac{4 (\epsilon_{eg} - \omega) m_i^2 }{3\epsilon_{eg}  (\epsilon_{eg} -2 \omega) }
\right)
\,.
\end{eqnarray}

\begin{figure*}[htbp]
 \begin{center}
 \epsfxsize=0.6\textwidth
 \centerline{\epsfbox{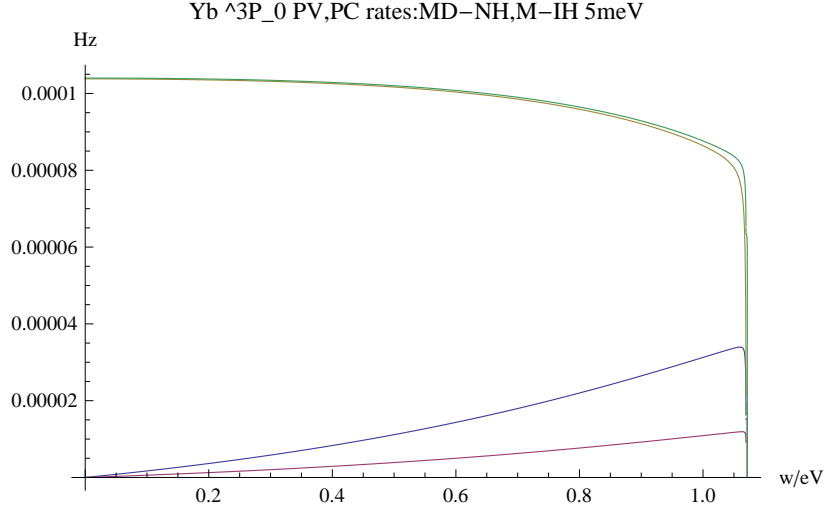}} \hspace*{\fill}
   \caption{${}^{171}$Yb$\,{}^3P_0$ RENP PC rates, 
and PV rate differences  under field reversal 
for Majorana neutrino pair emission of smallest mass 5meV,
   PV-NH in blue and PV-IH in  magenda,
   PC-NH in brown and PC-IH in green.
   PC rates are scaled down by 1/500 for easy comparison.
Assumed parameters:
   target number density $10^{22}$cm$^{-3}$,
   target volume $10^2$cm$^3$.
}
   \label{pvc 3p0 mnih overall}
 \end{center} 
\end{figure*}

Rates for ${}^3P_0$ RENP are illustrated  
in Fig(\ref{pvc 3p0 mnih overall})
by taking field reversal at $\pi/4$ and $3\pi/4$ angles.
Rates for ${}^3P_0$ RENP are typically smaller by an order of magnitudes
than ${}^3P_2$ RENP.

\vspace{0.5cm}
\lromn8
{\bf Summary}

\vspace{0.3cm}
We examined how parity violating
asymmetry and PV rate differences in RENP may be observed in atomic
de-excitation.
Large PV interference and PV asymmetry
may occur in transitions among different parity states,
which suggests alkaline earth atoms as good targets.
We have demonstrated that large $1/m_e$ suppression
inherent in non-relativistic electrons in atoms
are avoided by using the combined interference of
the nuclear mono-pole and the valence spin
pair emission.
Fundamental formulas applicable when
magnetic sub-levels are energetically resolved
are derived and used for numerical computations.
The PV asymmetry may reach of order a few times $10^{-3}$
in the examined case of Yb.
A further systematic search for better target atoms,
especially for ions
implanted in transparent crystals
is indispensable for realistic RENP experiments
along with extensive numerical simulations of
the dynamical factor $\eta_{\omega}(t)$.

\vspace{0.5cm}
\lromn9 
{\bf Appendices }

\vspace{0.3cm}
{\bf A: Weak hamiltonian of neutrino pair emission
}

\vspace{0.3cm}
We shall use the somewhat unfamiliar representation of
Clifford algebra, namely the representation of
diagonal $\gamma_5 = i \gamma_0 \gamma_1 \gamma_2 \gamma_3$  
(for the purpose of the clearest distinction of Dirac and Majorana 
fermions).
Hence it might be appropriate to clarify points that
might cause confusion to the reader,
which we shall explain following \cite{my-prd-07} 
(thereby correcting a formula there).

The basic weak hamiltonian density of neutrino pair emission
that appears in atomic transitions is given by 
\begin{eqnarray}
&& 
\frac{G_F}{\sqrt{2}}\,\left(
\bar{\nu}^e \gamma_{\alpha}(1 - \gamma_5)\nu^e
\bar{\psi}\gamma^{\alpha}(1 - \gamma_5) \psi
- \frac{1}{2}
\sum_i \bar{\nu}^i \gamma_{\alpha}(1 - \gamma_5)\nu^i 
\bar{\psi} 
\gamma^{\alpha}
(1 - 4\sin^2 \theta_W - \gamma_5)  \psi \right)
\,,
\end{eqnarray}
in the charge-retention order after Fierz transformation.

In atoms one may use expansion of the electron
field operator in terms of bound state and modified
plane wave functions, and we may safely ignore
contribution from the plane wave part for
our application.
Thus we may assume that the field operator satisfies
the Dirac equation;
in the $\gamma_5-$diagonal representation,
\begin{eqnarray}
&&
\psi =
\left(
\begin{array}{c}
\varphi  \\
\chi
\end{array}
\right)
\,,\hspace{0.5cm}
\left(
E - i\vec{\sigma}\cdot\vec{\nabla} - V
\right)\,\varphi = -m_e \chi
\,, \hspace{0.5cm}
\left(
E + i\vec{\sigma}\cdot\vec{\nabla} - V
\right)\, \chi
= -m_e \varphi
\,, 
\\ &&
\left(
E^2 + \vec{\nabla}^2
- 2E V - i \vec{\sigma}\cdot\vec{\nabla}V + V^2
- m_e^2
\right)\,\varphi
= 0
\,, \hspace{0.5cm}
\chi = - \frac{1}{m_e}\left(
E - i\vec{\sigma}\cdot\vec{\nabla} - V
\right)\,\varphi
\,.
\end{eqnarray}
The potential $V$ includes the Coulomb potential
and any other one-body correction to that, 
in the covariant $\gamma_0 V$ form.
In this $\gamma_5-$diagonal Majorana representation
two 2-component fields $\chi\,, \varphi$ are of
the same order even for non-relativistic electrons.

Four vector and axial vector currents can be written
in terms of two component spinors $\varphi$ of
electron field operator as follows.
Writing temporal and spatial components separately,
they are
\begin{eqnarray}
&&
\hspace*{-1cm}
\bar{\psi}' \gamma_{\alpha}\psi
= \varphi'^{\dagger}
\left(
1 + 
\frac{(E' - V + i\vec{\sigma}\cdot\vec{\nabla}')
(E - V - i\vec{\sigma}\cdot\vec{\nabla})}{m_e^2}
\right)\varphi \,,
\nonumber \\ &&
- \varphi'^{\dagger}
\left(
\vec{\sigma} - 
\frac{(E' - V + i\vec{\sigma}\cdot\vec{\nabla}')\vec{\sigma}
(E - V - i\vec{\sigma}\cdot\vec{\nabla})}{m_e^2}
\right)\varphi
\,,
\\ &&
\hspace*{-1cm}
\bar{\psi}' \gamma_{\alpha}\gamma_5 \psi
= \varphi'^{\dagger}
\left(
- 1 + 
\frac{(E' - V + i\vec{\sigma}\cdot\vec{\nabla}')
(E - V - i\vec{\sigma}\cdot\vec{\nabla})}{m_e^2}
\right)\varphi
\,,
\nonumber \\ &&
\varphi'^{\dagger}
\left(
 \vec{\sigma} + 
\frac{(E' - V + i\vec{\sigma}\cdot\vec{\nabla}')\vec{\sigma}
(E - V - i\vec{\sigma}\cdot\vec{\nabla})}{m_e^2}
\right)\varphi
\,,
\\ &&
\bar{\psi}' \gamma_{\alpha}( 1- \gamma_5) \psi
= 2
\left(
\varphi'^{\dagger}\varphi \,,
- \varphi'^{\dagger}\vec{\sigma}\varphi
\right) 
\,,
\end{eqnarray}
to all orders $1/m_e$.
$\vec{\nabla}'$ acting on functions to the left.

To extract $1/m_e$ terms, it is necessary to
subtract the rest mass energy $m_e$ from $E$ and
write $E= m_e+\epsilon$.
This procedure gives
\begin{eqnarray}
&&
{\cal H}_W = 
\frac{G_F}{\sqrt{2}} \sum_{ij} j_{ij}^{\alpha}j_{ij\,,\alpha}^e
\,, \hspace{0.5cm}
j_{ij}^{\alpha}= \nu_i^{\dagger} \sigma^{\alpha}\nu_j
\,,
\\ &&
\hspace*{-1cm}
j_{ij\,,0}^e =  e^{\dagger}
\left(
b_{ij} -2\sin^2 \theta_W \delta_{ij}
\vec{\sigma}\cdot\frac{ i\vec{\nabla}}{m_e}
\right)e
\,, \hspace{0.5cm}
b_{ij} = U_{ei}^*U_{ej} - 
\frac{\delta_{ij}}{2} (1 - 4\sin^2 \theta_W )
\,,
\\ &&
\hspace*{-1cm}
\vec{j}_{ij}^e = e^{\dagger} \left( a_{ij}\vec{\sigma}
 + 2\delta_{ij}\sin^2 \theta_W\frac{-i \vec{\nabla} - \vec{\sigma} 
\times \vec{\nabla}}{m_e}
\right)e
\,, \hspace{0.5cm}
a_{ij} = - U_{ei}^*U_{ej} + \frac{1}{2}\delta_{ij}
\,,
\label{weak 4-fermi}
\end{eqnarray}
with $\sigma^{\alpha} = (1\,, \vec{\sigma})$.
In writing this we changed the normalization factor
of two component spinors,
using the relation
between 4-component and 2-component wave functions,
\begin{eqnarray}
&&
\int d^3 x \bar{\psi_n} \psi_n
= -2 \int d^3 x \varphi_n^{\dagger}
\left( 1 - \frac{\epsilon_n + V}{m_e}
\right) \varphi_n
\,.
\end{eqnarray}
Except the factor 2, there is
a sign change between the two.
In the non-relativistic limit,
$(\epsilon_n + V)/m_e$ term is of order
$\alpha^2$ and small.

We point out the origin of non-relativistic
electron operators in Lorentz covariant currents
prior to taking the non-relativistic limit:
except the piece of term $\propto a_{ij}$ all other terms
come from 4-vector $V_{\alpha} \propto \gamma_{\alpha}$,
while the term $\propto a_{ij}$ arise from
4-axial vector $A_{\alpha} \propto \gamma_{\alpha}\gamma_5$.
Hence PV quantities arising from electron contributions
are all suppressed by $v/c \propto 1/m_e$ effect
except the interference contribution with the term $\propto b_{ij}$
arising from core electrons.
This favors heavy atoms since electron velocity
in heave atoms may be enhanced by some power of
atomic number $Z$.
The proposed PV quantity in the text is much
enhanced due to a novel interference
between the electron and the nuclear mono-pole
current, which does not suffer from 1/mass suppression
at all.

\vspace{0.5cm}
{\bf Appendix B: Neutrino phase space integral}

\vspace{0.3cm}
Using the helicity summation formula of \cite{my-prd-07}
and disregarding irrelevant T-odd terms,
one has 
\begin{eqnarray}
&&
\hspace*{1cm}
\sum_{h_i} |j_0^{\nu}\cdot A_0 
+ \vec{j}^{\nu}\cdot\vec{A} |^2 = 
\nonumber \\ &&
\hspace*{-1cm}
\frac{1}{2} (1 + \frac{\vec{p}_1\cdot\vec{p}_2}{E_1E_2}+\delta_M 
\frac{m_1 m_2}{E_1E_2} ) 
|A_0|^2
+ 
\frac{1}{2} (1 - \frac{\vec{p}_1\cdot\vec{p}_2}{E_1E_2} -\delta_M 
\frac{m_1 m_2}{E_1E_2} ) 
|\vec{A}|^2 +  \frac{\Re(\vec{p}_1\cdot\vec{A} \vec{p}_2\cdot\vec{A}^*)}{E_1 E_2}
-2 (\frac{  \vec{p}_1}{ E_1} + \frac{  \vec{p}_2}{ E_2})  \Re (A_0 \vec{A}^*)
\,,
\nonumber \\ &&
\label{helicity summation of currents}
\end{eqnarray}
where $(E_i, \vec{p}_i)$ are neutrino 4-momenta.
In the phase space integral of neutrino momenta,
\begin{eqnarray}
&&
\int d{\cal P}_{\nu}(\cdots)
= 
\int \frac{d^3 p_1 d^3 p_2}{(2\pi)^2} \delta(E_1 + E_2 + \omega - \epsilon_{eg}) 
\delta(\vec{p}_1 + \vec{p}_2 + \vec{k}) (\cdots)
\end{eqnarray}
one of the momentum integration is used to eliminate the
delta function of the momentum conservation.
The resulting energy-conservation is used to fix the relative angle
factor $\cos \theta $
between the photon and the remaining neutrino momenta, 
$\vec{p}_1 \cdot \vec{k} = p_1 \omega \cos \theta$.
Noting the Jacobian factor $E_2/p\omega$
from the variable change to the cosine angle, one obtains one dimensional integral
over the neutrino energy $E_1$:
\begin{eqnarray}
&&
\frac{1 }{2\pi \omega}
\int_{E_-}^{E_+} d E_1E_1 E_2 \frac{1}{2} (\cdots)
\,, \hspace{0.5cm}
E_2 = \epsilon_{eg} - \omega - E_1
\,.
\end{eqnarray}
The angle factor constraint $|\cos \theta| \leq 1$
places a constraint on the range of neutrino energy integration,
\begin{eqnarray}
&&
E_{\pm} =
\frac{1}{2} \left( (\epsilon_{eg} - \omega) (1 +
\frac{m_i^2 - m_j^2}{\epsilon_{eg}(\epsilon_{eg} - 2\omega)} )
\pm \omega \Delta_{ij}(\omega)
\right)
\,,
\\ &&
\Delta_{ij}(\omega) 
= \left\{
\left(1 - \frac{ (m_i + m_j)^2}{\epsilon_{eg} (\epsilon_{eg} -2\omega) } \right)
\left(1 - \frac{ (m_i - m_j)^2}{\epsilon_{eg} (\epsilon_{eg} -2\omega) } \right)
\right\}^{1/2}
\,.
\end{eqnarray}

We record for completeness all four important integrals over the neutrino pair momenta:
\begin{eqnarray}
&&
\int d{\cal P}_{\nu} \frac{ 1}{ E_1E_2} = \frac{\Delta_{12}(\omega)}{2\pi} 
\equiv D(\omega)
\,,
\\ &&
\hspace*{-1cm}
\int d{\cal P}_{\nu} 1 = \frac{\Delta_{12}(\omega)}{2\pi}
\left(
\frac{1}{4} (\epsilon_{eg} - \omega)^2 - \frac{\omega^2 }{12}
 +\frac{\omega^2 (m_1^2 + m_2^2)}{6\epsilon_{eg}  (\epsilon_{eg} -2 \omega) }
- \frac{\omega^2 (m_1^2 - m_2^2)^2}{12 \epsilon_{eg}^2  (\epsilon_{eg} -2 \omega)^2} 
- \frac{ (\epsilon_{eg} - \omega)^2 (m_1^2 - m_2^2)^2}{2 \epsilon_{eg}^2  (\epsilon_{eg} -2 \omega)^2} 
\right)
\equiv C(\omega)
\,,
\nonumber \\ &&
\\ &&
\hspace*{-1cm}
\int d{\cal P}_{\nu} ( \frac{\vec{p}_1}{ E_1} + \frac{\vec{p}_2}{ E_2}) = 
- \frac{\Delta_{12}(\omega)}{4\pi}\vec{k}
\left(
\epsilon_{eg} -\frac{4}{3} \omega
 +\frac{2 (\epsilon_{eg} - \omega) (m_1^2 + m_2^2)}{3\epsilon_{eg}  (\epsilon_{eg} -2 \omega) }
-\frac{4}{3} \frac{ (\epsilon_{eg} - \omega) (m_1^2 - m_2^2)^2}{ \epsilon_{eg}^2  (\epsilon_{eg} -2 \omega)^2} 
\right)
\equiv \vec{k}  \frac{J(\omega)}{\omega}
\,,
\label{vector integral}
\\ &&
\int d{\cal P}_{\nu} \frac{p_1^i  p_2^j +p_1^j p_2^i  }{2 E_1E_2} =
\frac{1}{2 } ( \delta_{ij} - \frac{k^i k^j}{\omega^2 } ) A(\omega)
 + \frac{1}{2 \omega^2 } (3\frac{k^i k^j}{\omega^2 } - \delta_{ij})B(\omega)
 \,,
\\ &&
A(\omega) =\int d{\cal P}_{\nu} \frac{\vec{p}_1\cdot \vec{p}_2}{E_1 E_2}
\nonumber \\ &&
\hspace*{-1cm}
=  \frac{\Delta_{12}(\omega)}{2\pi} 
\left( 
- \frac{1}{4}  (\epsilon_{eg} - \omega)^2
+ \frac{5}{12} \omega^2 +  \frac{1}{2} (m_1^2 + m_2^2 )
 +\frac{\omega^2 (m_1^2 + m_2^2)}{6\epsilon_{eg}  (\epsilon_{eg} -2 \omega) }
- \frac{1}{12}  \frac{(m_1^2 - m_2^2)^2}{\epsilon_{eg}^2  (\epsilon_{eg} -2 \omega)^2 }  
(\omega^2 + 3  (\epsilon_{eg} - \omega)^2\,)  
\right)
\,,
\\ &&
B(\omega) = \int d{\cal P}_{\nu} \frac{ \vec{k} \cdot\vec{p}_1 \vec{k} \cdot \vec{p}_2}{E_1 E_2}
= - \frac{\Delta_{12}(\omega)}{2\pi} 
\frac{\omega^2}{12} (\epsilon_{eg}^2 - 2 \omega\epsilon_{eg} - 2\omega^2 )
\,.
\end{eqnarray}

\vspace{0.5cm}
{\bf Appendix C: Intermediate coupling scheme in heavy atoms}

\vspace{0.3cm}
The spin orbit interaction given by
$\sum_i^2 \xi(r_i) \vec{l}_i \cdot \vec{s}_i$
causes energy shifts and  mixing of states
in the $LS$ coupling scheme.
These effects becomes more important in heavier atoms.
We shall derive energy eigenstates for
two-electron system of $ns, n'l$ (the angular momentua are $0, l$)
that includes
heavy alkaline atoms.
Our method is elementary and calculation
is straightforward.
More general method that can deal with more
complicated multi-electron system
is given in \cite{condon-shortley}.

There are four independent states with
the magnetic degeneracy of $2J+1$.
In the $LS$ basis they are
${}^3L_{l+1}, {}^3L_l, {}^1L_1, {}^3L_{l-1}$ with
$L = l$; three spin-triplet states and one
spin-singlet state.
Due to the angular momentum conservation
of the spin-orbit interaction the mixing occurs
between ${}^3L_l$ and ${}^1L_l$, and other states
become energy shifted.
The $LS$ basis we need in the following calculations
are decomposition into the direct product of
the spin and the orbital parts:
\begin{eqnarray}
&&
|{}^3P_{l},l \rangle  =\frac{1}{\sqrt{l+1}}
(   -|l, l-1\rangle_L^3 |1,1 \rangle_S + \sqrt{l}|l,l\rangle_L^3 
|1,0 \rangle_S )
\,,  
\\ &&
|{}^lP_l,l \rangle  = |l,l\rangle_L^1|0,0 \rangle_S 
\,, \hspace{0.5cm} 
|{}^3P_{l+1},l+1 \rangle  = |l,l\rangle_L^3|1,1 \rangle_S 
\,,
\\ &&
\hspace*{-0.5cm}
|{}^3P_{l-1},l-1 \rangle  =\frac{1}{\sqrt{l(2l+1)}}
( |l,l-2\rangle_L^3 |11 \rangle_S 
- \sqrt{2l-1} |l, l-1 \rangle_L^3 |1,0\rangle_S
+  \sqrt{l(2l-1)}|l,l\rangle_L^3 |1,-1 \rangle_S)
\,,
\\ &&
|l, m \rangle_L^3 = \frac{1}{\sqrt{2}}
(|l,m \rangle_1 |0,0\rangle_2 - |0,0 \rangle_1 |l,m\rangle_2)
\,, \hspace{0.5cm}
|l, m \rangle_L^1 = \frac{1}{\sqrt{2}}
(|l,m \rangle_1 |0,0\rangle_2 + |0,0 \rangle_1 |l,m\rangle_2)
\,.
\end{eqnarray}
We define the strength of the spin-orbit interaction
in terms of the single electron matrix element,
\begin{eqnarray}
&&
\langle nl jm | \xi(r) \vec{l}\cdot \vec{s} | nljm \rangle
= \zeta_{nl} \frac{1}{2} \left( j(j+1) - l(l+1) - \frac{3}{4}
\right)
\,.
\end{eqnarray}

We illustrate calculation of spin-orbit matrix
elements in an example,
\begin{eqnarray}
&&
\langle {}^3L_l, l | \sum_i \xi(r_i) \vec{l}_i\cdot\vec{s}_i
| {}^1L_1, l \rangle
= \langle {}^3L_l, l | \sum_i \xi(r_i) \left(
\frac{1}{2}(\vec{l}_{i+}\cdot\vec{s}_{i_-} + \vec{l}_{i-}\cdot\vec{s}_{i_+})
+ \vec{l}_{iz}\cdot\vec{s}_{i_z} \right)
| {}^1L_l, l \rangle
\,.
\end{eqnarray}
We note that the operation 
$s_{i-} |0,0\rangle_S \propto |1,-1\rangle_S$,
hence $\vec{l}_{i+}\cdot\vec{s}_{i_-}| ^1L_l, l \rangle$ 
has no overlap with the initial state $|^3L_l,l \rangle$,
as seen in formulas of the direct product decomposition.
Furthermore, the operation $s_{iz} |0,0 \rangle_S \propto
|1,0 \rangle_S$ simplifies calculation.
Thus, a part of the spin-orbit matrix element is
calculated as
\begin{eqnarray}
&&
\langle {}^3L_l, l |\xi(1) l_{1-} s_{1+} 
| {}^1L_l, l\rangle
= \frac{\zeta}{\sqrt{2(l+1)}}\, {}^3
\!\langle l, l-1 | l_{1-} | l,l\rangle^3
= \frac{\zeta}{2}\sqrt{\frac{l}{l+1}}
\,,
\end{eqnarray}
which is also equal to 
$\langle {}^3L_l, l |\xi(2) l_{2-} s_{2+} | {}^1L_l, l\rangle$.
Similarly, one has
\begin{eqnarray}
&&
\langle {}^3L_l, l |\xi(1) l_{1z} s_{1z} | {}^1L_l, l\rangle
= \frac{\zeta}{2}\sqrt{\frac{l}{l+1}
}\, {}^3\!\langle l, l-1 | l_{1z} | l,l\rangle^3
= \frac{\zeta }{4}l\sqrt{\frac{l}{l+1}}
\,.
\end{eqnarray}
Adding all of these non-vanishing elements, one obtains
\begin{eqnarray}
&&
\langle {}^3L_l, l | \sum_i \xi(r_i) \vec{l}_i\cdot\vec{s}_i
| {}^1L_l, l \rangle
= \frac{\zeta}{2} \sqrt{l(l+1)}
\,.
\end{eqnarray}

Other matrix elements are calculated in similar fashions.
Adding electrostatic energies and a common value for
all four energy levels, 
the entire energy matrix is given by
\begin{eqnarray}
&&
\hspace*{-1cm}
(|{}^3P_2 \rangle, |{}^3P_1 \rangle, 
|{}^1P_1 \rangle, |{}^3P_0 \rangle )
\left(
\begin{array}{cccc}
F-G + \frac{\zeta}{2}l & 0 & 0&0 \\
0 & F-G -\frac{\zeta}{2} & \frac{\zeta}{2}\sqrt{l(l+1)} & 0\\
0 & \frac{\zeta}{2}\sqrt{l(l+1)} & F+G & 0\\
0 & 0 & 0 & F-G - \frac{\zeta}{2}(l+1)
\end{array}
\right)
\left(
\begin{array}{c}
|{}^3P_2 \rangle  \\
 |{}^3P_1 \rangle  \\
 |{}^1P_1 \rangle  \\
 |{}^3P_0 \rangle    
\end{array}
\right)
\,.
\nonumber \\ &&
\end{eqnarray}
Energy eigenvalues for the mixed $sp$ two-electron states are,
with diagonalization, given by
\begin{eqnarray}
&&
\epsilon({}^3P_2) = F - G + \frac{\zeta}{2}
\,, \hspace{0.5cm}
\epsilon({}^3P_0) = F-G - \zeta
\,,
\\ &&
\epsilon_{\pm} = F- G - \frac{\zeta}{4} \pm
\sqrt{(G + \frac{\zeta}{4})^2 + \frac{\zeta^2}{2}}
\,.
\end{eqnarray}
Eigenstates corresponding to these energy eigenvalues are
given in terms of unperturbed basis,
\begin{eqnarray}
&&
|{}^+P_1 \rangle = \cos \theta |{}^1P_1 \rangle 
+ \sin \theta |{}^3P_1 \rangle
\,,
\\ &&
|{}^-P_1 \rangle = \cos \theta |{}^3P_1 \rangle 
- \sin \theta |{}^1P_1 \rangle
\,,
\\ &&
\tan (2\theta ) = \frac{2\sqrt{2} \zeta}{4G + \zeta}
\,.
\end{eqnarray}

There are three parameters, $F,G, \zeta$,
in this mixing problem, and
there exist four data of energies for these states.
Thus, all three parameters are determined by experimental data
and there is a further consistency relation (prediction)
among four energies.
A convenient choice is
\begin{eqnarray}
&&
\zeta = \frac{2}{3} \left( \epsilon({}^3P_2) - \epsilon({}^3P_0)
\right)
\,,
\\ &&
F-G = \frac{1}{3} \left( 2\epsilon({}^3P_2) + \epsilon({}^3P_0)
\right)
\,,
\\ &&
F+G = \frac{1}{3} \frac{9\epsilon_+\epsilon_-
+ 4 (\epsilon({}^3P_2) - \epsilon({}^3P_0)\,)^2}
{\epsilon({}^3P_2) +2 \epsilon({}^3P_0)}
\,.
\end{eqnarray}
The consistency relation is given by
\begin{eqnarray}
&&
\frac{1}{2}\left( \epsilon_+ + \epsilon_-
\right)
=
\frac{1}{6}
\left( \epsilon({}^3P_2) + 2\epsilon({}^3P_0)
\right)
+
\frac{1}{6} \frac{9\epsilon_+ \epsilon_-
+ 4 (\epsilon({}^3P_2) - \epsilon({}^3P_0)\,)^2}
{\epsilon({}^3P_2) +2 \epsilon({}^3P_0)}
\label {consistency of level energies}
\end{eqnarray}

The mixing angle $\theta$
calculated from experimental data of Yb $6s6p$-system
is $\sim 0.16$.
Accuracy of the consistency relation (\ref{consistency of level energies}) 
is something like 2.669 (LHS) vs 2.676(RHS) and is excellent.

Following \cite{condon-shortley},
one may use a convenient set of
parameters for the energy level diagram;
\begin{eqnarray}
&&
\eta = \frac{\epsilon}{G\sqrt{1+\chi^2}}
\,, \hspace{0.5cm}
\frac{\chi}{1+\chi}
\,,
\label{ec variables}
\end{eqnarray}
with $\chi \equiv 3\zeta/(4G)$, 
for the energy and for the strength of
$LS-$interaction, respectively.
In the weak coupling limit $\eta_{\pm} \rightarrow \pm 1$ 
as $\chi \rightarrow 0$,
and $\eta_{\pm} \rightarrow 2/3, -4/3$ as $\chi \rightarrow \infty$.

\begin{figure*}[htbp]
 \begin{center}
 \epsfxsize=0.6\textwidth
 \centerline{\epsfbox{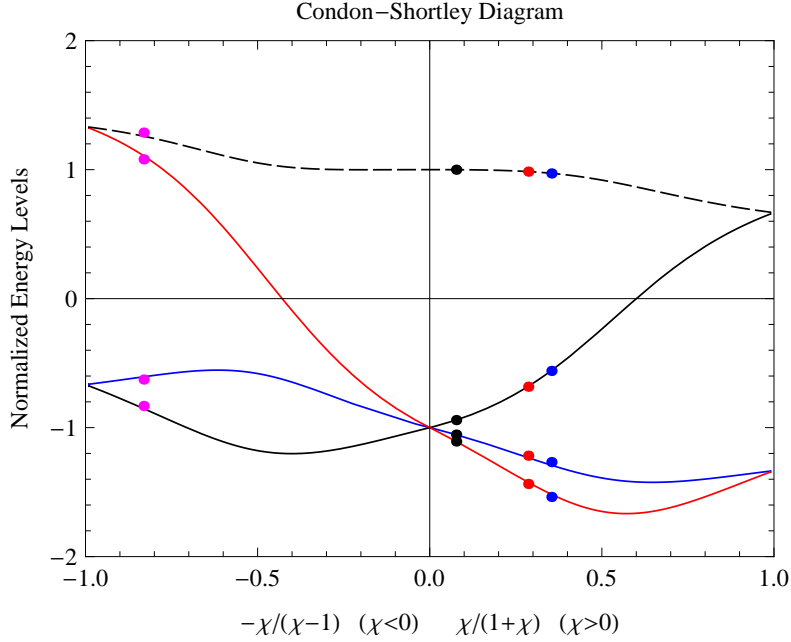}} \hspace*{\fill}
   \caption{Condon-Shortley diagram for alkali earth atoms:
   the variable $\eta$ plotted against $\chi/(\chi+1)$
of eq.(\ref{ec variables}).
Experimental data of level energies of 
${}^+P_1, {}^3P_2, {}^-P_1, {}^3P_0$
(from the upper to the lower)  of Sr, Yb, and Hg 
are plotted in dots; from the left to the right in the right $\chi> 0$.
Xe data of $\chi < 0$ are in the left.}
   \label{ec diagram}
 \end{center} 
\end{figure*}

The energy curves  are plotted in Fig(\ref{ec diagram}).
These curves are universal for all alkaline earth atoms.
Three sets of atomic data of Sr, Yb, and Hg and Xe of
electron-hole system of the same quantum numbers as alkaline
earth atoms are compared with these curves:
the agreement of theoretical curves and data are good.

\vspace{0.5cm}
{\bf Appendix D: Hyperfine interaction}

\vspace{0.3cm}
Nucleus may produce various types of
multi-pole fields.
Most important are magnetic field by
their magnetic dipole moment and
electric quadrupole field by nuclear quadrupole moment.

Magnetic hyperfine interaction arises from
interaction of electron magnetic dipole moment with
the nuclear magnetic field.
The interaction may be written in the product form
of electron and nuclear magnetic moments,
or their angular momenta.
A standard hyperfine interaction for a single
electron is \cite{weissbluth}
\begin{eqnarray}
&&
{\cal H}_h = g_e g_N \mu_B \mu_N \left(
\frac{\vec{I}\cdot (\vec{L}-\vec{S})}{r^3}
+ \frac{3\vec{I}\cdot\vec{r} \vec{S}\cdot\vec{r}}{r^5}
+ \frac{8\pi}{3} \delta(\vec{r})\vec{I}\cdot\vec{S} 
\right)
\,,
\label{magnetic hyperfine interaction}
\end{eqnarray}
where $\vec{S},\vec{L}$ are electron spin and orbital
angular momentum operator and $\vec{I}$ is the nuclear spin operator.
The first two terms are the dipole-dipole interaction
restricted to atomic orbitals of non-vanishing angular
momentum. while the third 
is the Fermi contact interaction restricted to s-orbitals
of non-vanishing wave functions at the origin of nucleus.

We have in mind applications in RENP,
and use of atoms with  two-electron system of two kinds;
alkaline earth like atoms made of $ns, n'p$.
Transition to intermediate state in RENP 
occurs by hyperfine interaction bridging different $J$ states
of two-electron, ${}^3P_2 \rightarrow {}^{\pm}P_1$
in alkaline earth atoms of odd isotopes such as ${}^{171}$Yb.

For simplicity we consider a specific transition
from the state of $F= M_F=3/2$.
Using the Wigner-Eckart theorem \cite{rose}, one can separate
the nuclear part and the atomic part as \cite{weissbluth}
\begin{eqnarray}
&&
\langle {}^{\pm}P_1 I J F | 
\vec{I}\cdot\vec{A}|{}^3P_2 I J' F  \rangle
= 
(-1)^{F +J' +I} 
\left\{
\begin{array}{ccc}
I & J'  & F \\
J & I & 1 
\end{array}
\right\}
\langle {}^{\pm}P_1 I ||\vec{I} || {}^{\pm}P_1 I \rangle
\langle  {}^{\pm}P_1 || \vec{A} || {}^3P_2 \rangle
\,,
\end{eqnarray}
with the 6j symbol here
\begin{eqnarray}
&&
\left\{
\begin{array}{ccc}
1/2 & J'  & 3/2 \\
J & 1/2 & 1 
\end{array}
\right\}
\,,
\end{eqnarray}
readily available for $(J', J) = (2,1)$.
Reduced matrix elements 
$\langle  {}^{\pm}P_1 || \vec{A} || {}^3P_2 \rangle$
are calculated using wave functions of two-electron system.
Final results may be expressed in terms of matrix elements of
single electron states;
\begin{eqnarray}
&&
a = g_e g_N \mu_B \mu_N\langle 6p | \frac{1}{r^3} | 6p \rangle
\,, \hspace{0.5cm}
b = g_e g_N \mu_B \mu_N\frac{8\pi}{3}|\psi_{6s}(0)|^2
\,.
\end{eqnarray}

In order to determine parameters $a, b$ of
hyperfine interaction, we calculate
hyperfine split levels.
Splitting follows the rule for different $F$;
$A \left( F(F+1) - J(J+1) - I(I+1) \right)/2$ 
even under the presence of dipole-dipole interaction \cite{weissbluth}.
Diagonal matrix elements we need are calculated as
\begin{eqnarray}
&&
\langle {}^3P_2\, 5/2 |\vec{I}\cdot \vec{A} | {}^3P_2\, 5/2 \rangle = \frac{1}{2\sqrt{6}} (b + \frac{14}{5}a)
\,,
\\ &&
\langle {}^3P_1\, 3/2 |\vec{I}\cdot \vec{A} | {}^3P_1\, 3/2 \rangle = \frac{1}{8} (b +10 a)
\,,
\\ &&
\langle {}^1P_1\, 3/2 |\vec{I}\cdot \vec{A} | {}^1P_1\, 3/2 \rangle = \frac{1}{2}a
\,.
\end{eqnarray}
Taking into account of the spin-orbit interaction,
the magnitudes of splitting are then given by
\begin{eqnarray}
&&
\epsilon({}^3P_2, F= 5/2) - \epsilon({}^3P_2, F= 3/2)= 
\frac{5}{4\sqrt{6}} (b + \frac{14}{5}a)
\,,
\\ &&
\epsilon({}^- P_1, F= 3/2) - \epsilon({}^-P_1, F= 1/2)= 
\frac{3}{8} (b +10a) 
\cos^2 \theta
- \frac{3}{2\sqrt{2}} (b + \frac{13}{5}a) \sin \theta \cos \theta
+ \frac{3}{2}a \sin^2 \theta
\,,
\\ &&
\epsilon({}^+ P_1, F= 3/2) - \epsilon({}^+P_1, F= 1/2)= 
\frac{3}{2}a \cos^2 \theta
+ \frac{3}{2\sqrt{2}} (b + \frac{13}{5}a) \sin \theta \cos \theta
+ \frac{3}{8} (b + 10 a)  \sin^2 \theta
\,.
\end{eqnarray}
No splitting exists for $\epsilon({}^3P_0, F= 1/2)$.
Without the spin-orbit interaction $\theta=0$ and
${}^+P_1$ hyperfine splitting is purely given 
by dipole-dipole interaction
$\propto a$.

From experimental data we may determine
$b \sim 12.6$GHz, $a\sim  0.17$GHz and
$\theta \sim - 0.04$.
The agreement of the spin-orbit mixing $\theta$ with the analysis in
the preceding Appendix C is not good.
The contribution
of dipole-dipole interaction is however much smaller 
$a/b \sim 1/70$ than the Fermi contact
term, and it would be sufficient to neglect
the dipole-dipole interaction in hyperfine splitting.
Moreover, the estimate of the spin-orbit mixing $\theta$
in the preceding section 
appears more reliable.

Hyperfine interaction causes mixing of states defined in the intermediate
coupling such as ${}^3P_2$ and ${}^{\pm}P_1$.
Magnitudes of this coupling are given in the text.

\vspace{0.5cm}
{\bf Appendic E: Magnetic factors and angular distributions}

\vspace{0.3cm}
One needs to consider two amplitudes, parity even (PE) and parity odd (PO),
in order to induce PV effects.
PE amplitude consists of hyperfine interaction sandwiched in time
sequence between the nuclear mono-pole neutrino pair emission
and E1 photon emission from the valence line.
PO amplitude consists of M1 type pair emission followed by
E1 photon emission.

We apply a magnetic field directed by an angle $\theta_m$ away from
the propagation (also the trigger axis).
All states are classified by specifying the magnetic
quantum numbers along the magnetic field.
We assume that at least electronic states
are energetically resolved by the magnetic field.

Disregarding energy denominators and coupling factors,
PE matrix elements are given in the $\vec{F}-$basis 
($\vec{F} = \vec{J} + \vec{I}$ 
is the total angular momentum of atoms and nucleus) by
\begin{eqnarray}
&&
-\sqrt{3} \sum_{F,M} d^F_{M_F', M}d^{1/2}_{M_F, M\pm 1} 
(-1)^{1/2 - M \mp 1}
\left(
\begin{array}{ccc}
1/2 & 1 & F \\
-M\mp 1  & \pm 1 & M 
\end{array}
\right)
\\ &&
\hspace*{0.5cm}
\cdot
\langle {}^{3,1}P_1\, F=3/2, M_F' |\vec{I}\cdot\vec{A}|{}^3P_2\, F=3/2,
M_F' \rangle 
\,.
\label{hf element}
\end{eqnarray}
We may define the magnetic factor for PE amplitude
disregarding the angle independent factor of eq.(\ref{hf element})
and calculate ($\pm$ coresponding to circular polarizations of $\pm 1$)
\begin{eqnarray}
&&
\sum_{F,M} d^F_{M_F', M}d^{1/2}_{M_F, M\pm 1} 
(-1)^{1/2 - M \mp 1}
\left(
\begin{array}{ccc}
1/2 & 1 & F \\
-M\mp 1  & \pm 1 & M 
\end{array}
\right) 
\nonumber \\ &&
= \frac{1}{2\sqrt{3}} (d^{3/2}_{M_F',-1/2}d^{1/2}_{M_F,1/2} 
+ \sqrt{3}d^{3/2}_{M_F',-3/2}d^{1/2}_{M_F,-1/2}) 
= \frac{1}{2} d^1_{1, -1}
\,,
\hspace{0.3cm} {\rm for}\; h= 1
\\ &&
= \frac{1}{2\sqrt{3}} (d^{3/2}_{M_F',1/2}d^{1/2}_{M_F,-1/2} 
+ \sqrt{3}d^{3/2}_{M_F',3/2}d^{1/2}_{M_F,1/2})
= \frac{1}{2} d^1_{1, 1}
\,,
\hspace{0.3cm} {\rm for}\; h= -1
\end{eqnarray}
The equality of d-functions is as expected since
one can equally work out the magnetic factor
without the nuclear spin in this case.

The PO amplitude has matrix element product in
the $\vec{J}-$basis,
\begin{eqnarray}
&&
\langle {}^1S_0| Y_{1,\pm 1}S_z |{}^3P_2, \tilde{M_J}' \rangle
\nonumber \\ &&
\hspace*{-1cm}
= \frac{1}{\sqrt{30}} \langle {}^1S_0|| \vec{Y}|| {}^{\pm}P_1\rangle
\langle {}^{\pm}P_1||\vec{S}|| {}^3P_2 \rangle
\sum_M
d^1_{M, \mp 1}
(- d^2_{M_J', 1}d^1_{M, 1} + \frac{2}{\sqrt{3}}d^2_{M_J', 0}d^1_{M, 0}
- d^2_{M_J', -1}d^1_{M, -1}) 
\,,
\\ &&
\langle {}^{+}P_1||\vec{S}|| {}^3P_2 \rangle = \sqrt{\frac{5}{2}}
\sin \theta
\,,\hspace{0.5cm}
\langle {}^{-}P_1||\vec{S}|| {}^3P_2 \rangle = \sqrt{\frac{5}{2}}
\cos \theta
\,,\hspace{0.5cm}
\langle {}^1S_0|| \vec{Y}|| {}^{\pm}P_1\rangle 
= - \sqrt{3}
\,.
\end{eqnarray}

In order to define magnetic factors, we introduce
\begin{eqnarray}
&&
W_2^{\pm}(x) = \sum_M d^1_{M, \pm 1}(x)
\left( -\sqrt{3} d^2_{1,1}(x) d^1_{M, 1}(x) + 2 d^2_{1,0}(x) d^1_{M, 0}(x)
-\sqrt{3} d^2_{1,-1}(x) d^1_{M, -1}(x)
\right)
\nonumber \\ &&
W_2^+(x) = \frac{\sqrt{3}}{2}  \left( -\cos x + \cos (2x) \right)
\,, \hspace{0.5cm}
W_2^-(x) = \frac{\sqrt{3}}{2} \left( -\cos x - \cos (2x) \right)
\,,
\\ &&
W_1^{+}(x) = d^1_{1,- 1}(x) = \sin^2 \frac{x}{2}
\,, \hspace{0.5cm}
W_1^{-}(x) = d^1_{1, 1}(x) =  \cos^2 \frac{x}{2}
\,.
\end{eqnarray}
Explicit functional forms have been calculated using
Mathematica.

Adding two circular polarizations gives
the angular distribution from a polarized atom,
for instance from $J=2, M_J=1$ in the example above.
These are given by elementary functions:

PV quantity
\begin{eqnarray}
&&
W_1^+W_2^+ + W_1^-W_2^- = -\sqrt{3} \cos^3 x
\,.
\end{eqnarray}

PC quantities
\begin{eqnarray}
&&
(1) \; 
(W_1^+)^2 + (W_1^-)^2 = \frac{1}{4}(3 + \cos (2x)\,)
\,,
\\ &&
(2) \; 
(W_2^+)^2 + (W_2^-)^2 = \frac{3}{4}(2 + \cos (2x) + \cos(4x) \,)
\,.
\end{eqnarray}
These functions satisfy the obvious constraint
under the transformation, $x \rightarrow \pi - x$.
PC quantities must be even under this transformation,
while PV quantity may contain odd piece.

\begin{figure*}[htbp]
 \begin{center}
 \epsfxsize=0.6\textwidth
 \centerline{\epsfbox{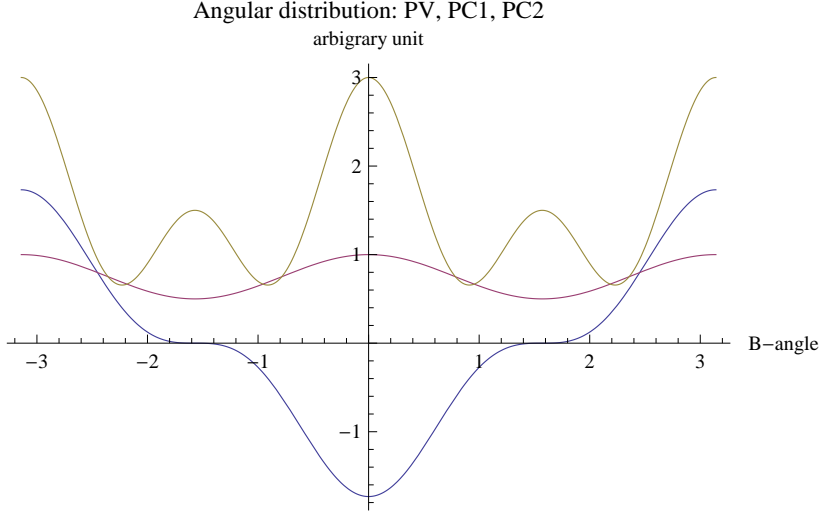}} \hspace*{\fill}
   \caption{Angular distributions for PV rate difference in blue, two PC rates
in magenda and brown,
the angle measured from the magnetic field direction
   for ${}^3P_2\, M_J=3/2 \rightarrow {}^1S_0\, 1/2$
}
   \label{pvc-angular-dist}
 \end{center} 
\end{figure*}

In Fig(\ref{pvc-angular-dist})
the magnetic field directional dependence
as given by the formulas above
is illustrated for the transition $F=3/2,M_F = 3/2 \rightarrow 
F=1/2, M_F= 1/2$.
Two contributions of different circular polarizations are added.

\vspace{0.5cm}
{\bf Acknowledgements}
\hspace{0.2cm}
We should like to thank T. Wakabayashi for a valuable discussion.
This research was partially supported by Grant-in-Aid for Scientific
Research on Innovative Areas "Extreme quantum world opened up by atoms"
(21104002)
from the Ministry of Education, Culture, Sports, Science, and Technology.


\begin{thebibliography}{99}
\bibitem{eta in ptep overview}
In \cite{ptep overview} a result for
numerical simulation of $\eta_{\omega}(t)$
is presented for pH$_2$ molecule target
(strong source of paired super-radiance (PSR)
of E1 $\times$ E1 transition, and  see
Fig 14 of this reference for time dependence).
Its time dependence is complicated:
a fast rise in $O(2$ ns), then a plateau region
of magnitude $O(10^{-2} \sim 10^{-3})$ of duration of
several nano-seconds, finally gradual decrease
ending around $10^{-6}$ at $\sim$ 12 ns (end time of calculation).
For RENP rate calculations,
numerical simulations based on the master
equation given in \cite{ptep overview} should be
performed for weaker PSR process of 
specific targets considered,
which is expected to give different time
profile and larger values of $\eta_{\omega}(t)$.
\end{thebibliography}
\end{document}